\input{aipcheck}

\documentclass[
    ,final            % use final for the camera ready runs
  ]
  {aipproc}
\usepackage{youngtab}
\layoutstyle{8x11single}
\def\ohalf{{\textstyle{1\over 2}}}
\def\half{{\textstyle{1\over 2}}}

\def\osix{{\textstyle{1\over 6}}}

\def\thalf{{\textstyle{3\over 2}}}

\def\M{{\cal M}}
\def\P{{\sf P}}

\def\D{{\cal D}}
\def\I{{\cal I}}

\newcommand{\beq}{\begin{equation}}
\newcommand{\be}{\begin{equation}}
\newcommand{\eeq}{\end{equation}}
\newcommand{\ee}{\end{equation}}
\newcommand{\beqa}{\begin{eqnarray}}
\newcommand{\bea}{\begin{eqnarray}}
\newcommand{\eeqa}{\end{eqnarray}}
\newcommand{\eea}{\end{eqnarray}}
\newcommand{\bra}[1]{\langle {#1} |}                        % < |
\newcommand{\ket}[1]{| {#1} \rangle}

\Yvcentermath{1}
\begin{document}

\title{Phenomenological hadron form factors: shape and relativity}


\author{B. Juli\'a-D\'{\i}az}
{address={Departament de Estructura i Constituents de la Materia,\\
Diagonal 647, Universitat de Barcelona, 08028 Spain}}

\begin{abstract}
The use of relativistic quark models with simple parametric 
wave functions for the understanding of the electromagnetic 
structure of nucleons together with their electromagnetic 
transition to resonances is discussed. The implications of 
relativity in the different ways it can be implemented in a 
simple model are studied together with the role played by mixed 
symmetry s-state and D-state deformations of the rest frame wave 
functions of the nucleon and $\Delta$ resonance.
\end{abstract}

\maketitle

\section{Introduction}

Understanding the structure of hadrons from the underlying 
theory of the strong interaction, QCD, has proved to be a 
quite demanding task from the theoretical point of view. 
The most promising approach which claims to have more 
solid connections to the theory is lattice-QCD where 
the electromagnetic structure of hadrons is beginning 
to be unveiled as the computational power keeps increasing.

We are confronted during the last years with a very 
interesting situation: experiments keep improving precision 
on hadronic observables and also keep providing finer 
and finer databases for electromagnetic observables, 
both for elastic processes and for transitions. As of 
today, we are not able to compute those observables 
from QCD and are still far from being able to do so in 
a well grounded lattice-QCD computation. Thus it becomes 
necessary to resort to models which should enable us to 
get a partial understanding of the processes at hand and 
which should above all serve to guide experiments.

Quark models are an ideal tool for that purpose, they 
incorporate part of the symmetries and the degrees of 
freedom of the original problem and permit to analyze 
the importance of some of them, namely the relevance of 
relativity and of the different configurations in the rest 
frame wave functions. The use of quark models to study 
electromagnetic form factors of nucleons and transitions 
to resonances has some history to which we cannot do 
justice in these proceedings. The reader may check 
Refs.~\cite{Chung,simula,Eli,boffi,mauro, julia04} and 
references therein if he is interested in following the 
quark model path along these last years. Many of the 
fine details of the results shown here are given in 
Refs.~\cite{julia04,dstate}.

The contribution is organized as follows, in the next 
section a formal description of the quark model and of 
the way electromagnetic form factors are extracted from 
matrix elements of the electromagnetic current is given 
for the elastic case. Then in section III the elastic 
electromagnetic form factors are considered and explored 
in detail. Section IV is devoted to the transition to 
the $\Delta$ resonance, exploring the effect of D-state 
configuration on the rest frame wave functions, 
Section V contains a final summary and discussion.

\section{The baryon model}

A simple way to build a descriptive phenomenological model 
is by constructing a mass operator in its spectral 
representation, being the different states of the 
representation the physical states. The ground state wave 
functions are built with a set of parameters which can 
be fitted to reproduce part of the known experimental data. 
This approach, in a more rigorous rendition, is explained 
in Ref.~\cite{julia04}.

Considering the $N$, $\Delta$ and $N^*(1440)$, their wave 
functions are constructed in the SU(6) symmetric quark model. 
The $N$ is built as the ground state, the $\Delta$ as the 
spin-flip excitation of the ground state, and the $N^*(1440)$ 
as a radial excitation. This assumption is of course restrictive 
with respect to further components in the wave functions. The 
main improvement to these is the consideration of $qqqq\bar{q}$ 
components. This is out of the scope of this contribution, 
advances in this direction have been reported in 
Ref.~\cite{RiskaLi}.

Those symmetric components can be written as:
\beqa
\Psi_{N:s} &=& \yng(3)_X \; {\yng(1,1,1)}_c \; {\yng(3)}_{SF} \nonumber \\
\Psi_{\Delta:s} &=& \yng(3)_{\,X} \; {\yng(1,1,1)}_{\,c} \; 
{\yng(3)}_{\,S}\,{\yng(3)}_{\,F}\nonumber \\
\Psi_{N^*:s} &=& \yng(3)_{\,X}^* \; {\yng(1,1,1)}_{\,c} \; {\yng(3)}_{\,SF} \, ,
\eeqa
where $X$, $c$, $S$, and $F$ stand for spatial, color, spin and 
flavor respectively. Other symmetry components which we will 
consider are mixed symmetry s-state ones, written as:
\beq
\Psi_{N:ms} = {\yng(1,1,1)}_c \; 
{1\over 2}\left[\; 
{\yng(2,1)}_{X:s} \;{\yng(2,1)}_{FS:s} +  
{\yng(2,1)}_{X:a} \;{\yng(2,1)}_{FS:a} 
 \right] \, ,
\eeq
and $D-$state ones, which for the nucleon and the $\Delta$ read:
\beqa
\Psi_{N:Ds} &=& 
{\yng(1,1,1)}_c \; 
 {1\over \sqrt{2}} \sum_{m s} 
(2 \thalf m s| \ohalf j_3)
\left[ \; {\yng(2,1)}_{X:s}^{\ell=2}\,  {\yng(2,1)}_{F:ms} 
 +  
{\yng(2,1)}_{X:a}^{\ell=2}\, {\yng(2,1)}_{F:ma} \right] \yng(3)_{\,S}\nonumber\\
\Psi_{\Delta:Ds} &=& 
{\yng(1,1,1)}_c \; 
 {1\over \sqrt{2}} \sum_{m s} (2 \ohalf m s| \thalf j_3)
\left[ \;
{\yng(2,1)}_{X:s}^{\ell=2}, {\yng(2,1)}_{S:ms} 
 +  
{\yng(2,1)}_{X:a}^{\ell=2}\, {\yng(2,1)}_{S:ma} \right] \yng(3)_{\,F} \, .
\eeqa

\begin{figure}[t]
\includegraphics[width=0.45\textwidth]{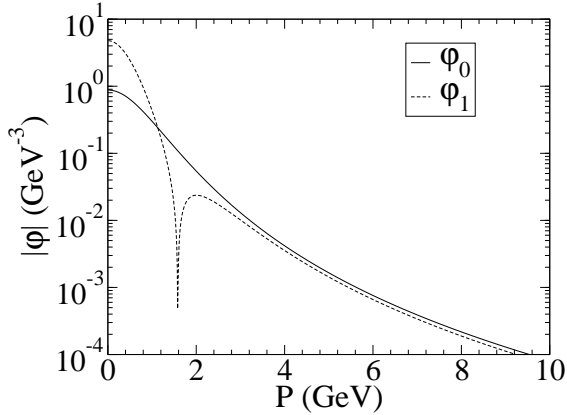}
\caption{Spatial wave functions.\label{fig:wf}}
\end{figure}

The spin-flavor components are written explicitly:
\beqa
{\yng(3)}_{FS} &=&
{1\over 2}\left[\; {\yng(2,1)}_{S:s}{\yng(2,1)}_{F:s} + 
{\yng(2,1)}_{S:a}{\yng(2,1)}_{F:a} \right]  \\
{\yng(2,1)}_{FS:s} &=&
{1\over 2}\left[\; {\yng(2,1)}_{S:s}{\yng(2,1)}_{F:s} - 
{\yng(2,1)}_{S:a}{\yng(2,1)}_{F:a} \right]\\
{\yng(2,1)}_{FS:a} &=&
{1\over 2}\left[\; {\yng(2,1)}_{S:s}{\yng(2,1)}_{F:a} + 
{\yng(2,1)}_{S:a}{\yng(2,1)}_{Fs:} \right]
\eeqa
and the spatial wave functions are written in momentum space as:
\beqa
\yng(3)_{\,X} &=&  {\cal N} 
\left( 1+ {\P^2\over 4 { b^2}}\right)^{-{ a}} \\
{\yng(2,1)}_{X:s} &=&  {\cal N}_s{\kappa^2-q^2\over\kappa^2
+q^2}\varphi_0(\kappa,q)\qquad 
{\yng(2,1)}_{X:a} =  {\cal N}_a{\vec\kappa\cdot\vec q\over\kappa^2
+q^2}\varphi_0(\kappa,q) \, ,
\eeqa
using the following momenta for the three quark system: 
$
\vec\kappa=\sqrt{2\over3} \left(\vec k_1 - {\vec k_2 + \vec k_3 \over 2}
\right), \;\vec q = {1\over \sqrt{2}} \left(\vec k_2 -\vec k_3 \right)
$, and 
$ \P:= \sqrt{2 (\vec\kappa^2+\vec q^2)}\, .$
The radial ground state wave function contains 2 parameters, 
which are fixed together with the constituent quark mass to 
reproduce the magnetic form factor of the proton in each of 
the forms of kinematics, which will be shown in the following 
section. The parameter values obtained for each of the forms 
are given in Table~\ref{tab:par}. 
\begin{table}[b]
\begin{tabular}{lcccc}
       & $m_q$ (MeV)  & $b$ (MeV) & $a$ &$r_0$ (fm) \\ 
\hline
point form   &  350           &  640        &  9/4 & 0.19 \\
front form   &  250           &  500        &   4  & 0.55 \\ 
instant form &  140           &  600        &   6  & 0.63  \
\end{tabular}
\caption{Parameter entering in the radial wave functions and 
light quark mass for each of the forms of kinematics.\label{tab:par}}
\end{table}
The spatial wave function of the first radial excitation is given 
explicitly in Ref.~\cite{julia04}. It is constructed from the 
ground state one by imposing both orthogonality and the presence 
of a node in a simple way. In Fig.~\ref{fig:wf} the ground 
state ($\varphi_0$) and first radial excitation ($\varphi_1$) 
are plotted. 

\subsection{Relativity and quark currents}

From a practical point of view, and due to the space 
constraints of these proceedings, let us explain in words 
the use of the different forms of relativistic kinematics. 
The main reason why relativity needs to be accounted for 
can be seen in a rough estimate of the quark velocity 
inside a proton. Once this is agreed there are in the 
literature three main ways of building relativity into 
a hamiltonian formalism, first discussed by Dirac~\cite{dirac} 
and later well developed in Refs.~\cite{polyzou,coe92}. In 
a non rigorous way which is well suited for explaining our 
procedure let us present the following rendition of the 
different forms. 

The three forms differ among themselves in the kinematical 
group of the Poincar\`e group, that is, the subgroup whose 
commutator relations are not affected by the interactions. 
According to this classification we have: instant form, 
where the subgroup is made of rotations and translations at 
a fixed time, point form, where boosts and rotations are 
kinematic, and front form, where, for example, boosts along 
the light cone are kinematical. 

For each form of kinematics the dynamics generates the 
current-density operator from a kinematic current, which 
is specified by the expression:
\be
\bra{\vec v_f,\vec v_2',\vec v_3'}I^\mu(0)
\ket{\vec v_3,\vec v_2,\vec v_a}=\delta^{(3)}(v_3'-v_3)
\delta^{(3)}(v_2'-v_2)\left(\osix+\half \tau_3^{(1)}\right)\bar u(\vec v_1\,')
\gamma^{(1)\mu}u(\vec v_1)\, ,
\label{cur4}
\ee
in the case of Lorentz kinematics, and:
\bea
&&\bra{P^+, P_{\perp f},{\bf p}_2',{\bf p_3}'}I^+(x^-,x_\perp)
 \ket{{\bf p}_3,{\bf p_2}, P_{\perp a},P^+}\cr
&&=\delta^{(3)}(p_3'-p_3)
\delta^{(3)}(p_2'-p_2)(\osix+\half \tau_3^{(1)})
\bar u({\bf p_1'}) \gamma^{(1)+}u({\bf p_1})
e^{\imath( P_{\perp f}- P_{\perp a})\cdot  x_\perp}\, ,
\eea
for light-front kinematics and finally by:
\bea
&&\bra{\half \vec  Q,\vec p'_2,\vec p_3'}I^\mu(\vec x)
 \ket{\vec p_3,\vec p_2,-\half \vec Q }=\delta^{(3)}(p_3'-p_3)
\delta^{(3)}(p_2'-p_2)(\osix+\half \tau_3^{(1)})\nonumber\\
&&\bar u(\vec p_1\,')\gamma^{(1)\mu}u(\vec p_1)
e^{\imath(\vec Q\cdot \vec x)}\, ,
\eea
for instant kinematics. $p_i$ and $p'_i$ are initial and 
final momenta of quark $i$. $v_i$ and $v'_i$ are initial 
and final velocities of quark $i$ respectively. In each 
case only covariance under the kinematic subgroup is required.

\begin{figure}[t]
\includegraphics[width=0.45\textwidth]{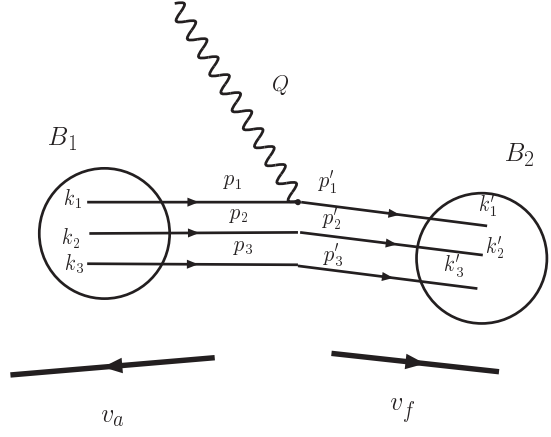}
\caption{Pictorial representation of the momentum variables 
at rest and in the moving frame.\label{fig:breit}}
\end{figure}

In Fig.~\ref{fig:breit} the role played by the form of 
kinematics which relates the variables at rest $\vec{k}_i$ 
and the ones at the vertex $\vec{p}_i$ is pictured graphically. 
These relations between both the rest frame spins and momenta 
to those of the moving frame depend on the form at use, their 
explicit formulae are given in Ref.~\cite{julia04}. Here it 
suffices to be aware that in those relations is where the 
relativistic nature of the computation enters.  

\section{Nucleon form factors}

The cross section for the elastic electron-nucleon scattering 
may be written in terms of two form factors. In the literature there 
are two usual sets, Pauli and Dirac form factors, $F_1$ and $F_2$ or 
Sachs form factors, $G_E$ and $G_M$. Both sets are related by 
kinematical factors. 

The elastic form factors can be extracted from the electromagnetic 
current taking appropriate matrix elements of spin states. They can 
be defined through the following matrix elements of the current 
operator for the case of instant and point form:
\bea
G_E(\eta)&:=&\bra{\half} \I_e(\eta)\ket{\half} _c= \bra{-\half} 
\I_e(\eta)\ket{-\half}_c\;  \nonumber\\
G_M(\eta)&:=&{1\over \sqrt{\eta}}  
\bra{\half}\I_{mx}(\eta)\ket{-\half}_c=-{1\over \sqrt{\eta}}  
\bra{-\half}\I_{mx}(\eta)\ket{\half}_c  \;. \label{GEGM}
\eea
with $\eta =Q^2/4M_N^2$. 

In the front form case, the form factors are extracted from 
matrix elements of the $+$ component of the current, defined 
as $I^+=I^0+I^z$:
\bea
F_1 &=& \langle \half | I^+ |\half\rangle \qquad 
F_2 = {1/\sqrt{\eta}} \langle -\half | I^+ |\half\rangle
\label{F1F2} \, .
\eea

As an explicit example, one can evaluate the matrix elements 
of Eq.~(\ref{GEGM}) with the antisymmetric nucleon wave function 
and the quark current (\ref{cur4}) multiplied by 3 (the number 
of constituent quarks). After summing over spin and isospin 
indices we arrive to the explicit expressions of the form factors 
in instant and point form:
\begin{eqnarray}
G_E(\eta)&=&\int d^3 p_2 d^3 p_3 \,
\varphi\left({{\kappa'}^2+q^{'2}\over 2 b^2}\right)
\varphi\left({\kappa^2+q^2\over 2 b^2}\right)
\sqrt{{\cal J}_{fa}(\vec p_2,\vec p_3)} 
C_{23}(\eta,\vec p_2,\vec p_3){\cal S}_e(\eta,\vec p_2,\vec p_3),
\cr\cr\cr
G_M(\eta)&=&\int d^3 p_2 d^3 p_3 
\varphi\left({{\kappa'}^2+q^{'2}\over 2 b^2}\right)
\varphi\left({\kappa^2+q^2\over 2 b^2}\right)
\sqrt{{\cal J}_{fa}(\vec p_2,\vec p_3)}
C_{23}(\eta,\vec p_2,\vec p_3){\cal S}_m(\eta,\vec p_2,\vec p_3)\; .\cr
&&
\label{GEM}
\end{eqnarray}
The Jacobian factor ${\cal J}_{fa}$,
\beq
{\cal J}_{fa} := J(v_f,\vec p_2,\vec p_3)
J(v_a,\vec p_2,\vec p_3)\, ,
\label{JAC}
\eeq
where $J$ are given explicitly in Ref.~\cite{julia04}. The 
jacobians differ in instant and point form.

The coefficient $C_{23}(\eta,\vec p_2,\vec p_3)$ is determined by 
the spectator Wigner rotations:
\bea
C_{23}(\eta,\vec p_2,\vec p_3)&=&{1\over 2} \sum_{\sigma',\sigma}
\left[\sum _{ \sigma''}
{\D^{1/2}_{\sigma' \sigma''}}^\dagger\left({\cal R}_W[B(v_{Kf}), k'_2]\right)
\D^{1/2}_{\sigma'',\sigma}\left({\cal R}_W[B(v_{Ka}), k_2]\right)\right]\cr\cr
&\times&
\left[ \sum_{\sigma''}{\D^{1/2}_{-\sigma' \sigma''}}^\dagger\left({\cal R}_W[B(v_{Kf}), k'_3]\right)
D^{1/2}_{\sigma'',-\sigma}\left({\cal R}_W[B(v_{Ka}), k_3]\right)\right]\; .
\label{C23}
\eea
The boost velocities $\vec v_{Ka}$, $\vec v_{Kf}$ are $\vec v_a$, $\vec v_f$ 
in point form and $\half \vec Q/\M_0'$, $-\half \vec Q/\M_0$ with instant 
kinematics. $\M_0$ is defined as $\M_0^2= (\sum _i E_i)^2-|\vec P|^2$.
The corresponding expressions for front form kinematics can be obtained using 
the explicit Melosh rotations given in Ref.~\cite{julia04}.

\subsection{Numerical results}

The three parameters of the model in each of the forms of 
kinematics are fitted to achieve a good reproduction of both 
the $Q^2$ dependence of the magnetic form factor of the proton 
and of its magnetic moment. The three forms accommodate the 
experimental data for the form factor and the pursued 
agreement is found with each of them. This was not guaranteed 
as we cannot assess apriori the importance of higher 
contributions, such as exchange currents, which may in each 
of the forms have a different relative importance. This 
happens for example when studying the form factor of a 
quark-antiquark pair which are bound to form a low mass 
system, e.g. a pion, there the point form has been proved 
to be unable to get close to the data~\cite{desplanques,junheplb,junheepja}.

\begin{figure}[t!]
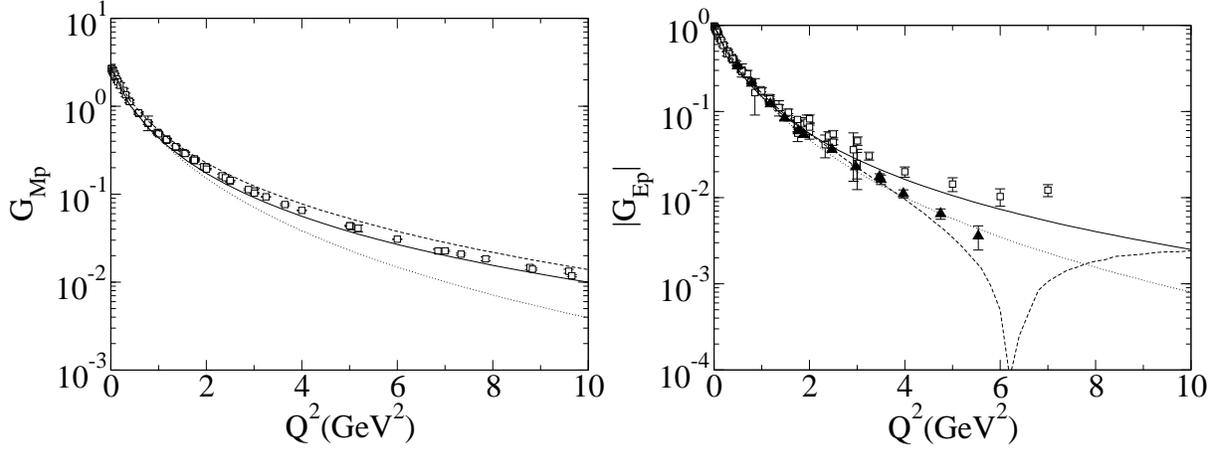

\vspace{20pt}
\includegraphics[width=0.48\textwidth]{fig_bjd_3.eps}
\includegraphics[width=0.48\textwidth]{fig_bjd_4.eps}
\caption{(left) Magnetic form factor of the proton in the symmetric quark 
model, solid, dotted and dashed stand for instant, point and front forms 
respectively. (right) Electric form factor of the proton. Solid, dotted and 
dashed lines correspond to the instant, point and front forms respectively. 
Squares are from the compilation of Ref.~\protect\cite{gep} while black 
triangles are obtained from the recent JLAB data of 
Refs.~\protect\cite{jones} using 
$G_{Ep} = (\mu_p G_{Ep} / G_{Mp})/(1+Q^2/0.71)^2$.
\label{fig:gmp}}
\end{figure}

The description of the magnetic form factor can be 
seen in Fig.~\ref{fig:gmp}. The low $Q^2$ domain is essentially 
well reproduced irrespective of the form at use. The high $Q^2$ 
data, on the contrary is better understood when instant and 
front form are considered, point form underestimates the 
experimental data above 3 GeV$^2$. This behavior at high $Q^2$ 
was already reported by the Graz group~\cite{boffi}. 

Once the parameters are fixed in each of the forms the remaining 
form factors of the nucleon were explored. The electric form factor 
of the proton is plotted in Fig.~\ref{fig:gmp}. At low $Q^2$ all 
forms reproduce the experimental data, giving an accurate 
description of the charge radius of the proton, given in 
Table~\ref{tab:zeroq}. This fall-off at low $Q^2$ is in fact very 
close to the magnetic form factor one, both approximately dropping 
as the standard dipole. The differences found in this form 
factor above 3 GeV$^2$ are already quite interesting as can be 
seen in the figure. First of all, an abrupt qualitative difference among 
the forms can be noticed, while point and instant forms remain 
positive up to $Q^2=10$ GeV$^2$ in front form the electric form 
factor of the proton becomes negative at around 6 GeV$^2$. This 
was already in the light-front computations of Chung and 
Coester~\cite{Chung} although there $F_1$ and $F_2$ are plotted 
instead of $G_E$ and $G_M$. This would be mostly anecdotical if 
it was not from the fact that the recent form factor data measured 
at JLAB using polarization transfer techniques exhibit a similar 
trend~\cite{jones}. The expected zero crossing appearing at 
$\approx Q^2=$ 7.5 GeV$^2$~\cite{arrington}.

\begin{figure}[t!]
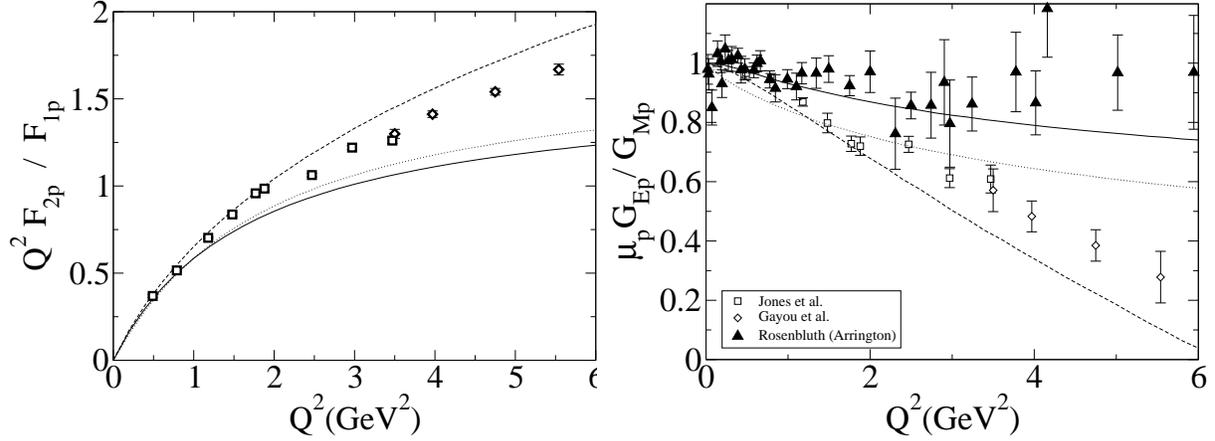

\vspace{20pt}
\includegraphics[width=0.48\textwidth]{fig_bjd_5.eps}
\includegraphics[width=0.48\textwidth]{fig_bjd_6.eps}
\caption{Relevant ratios, $G_{Ep}/G_{Mp}$ measured recently 
at JLAB and $Q^2 F_2/F_1$, description as in 
Fig.~\ref{fig:gmp}.\label{fig:ratios}}
\end{figure}

This $Q^2$ dependence of the electric form factor, the 
accessible quantity is in fact the ratio $G_E/G_M$, was 
not expected in the first pQCD predictions where it was 
believed that the ratio $Q^2 F_2 /F_1$ would be flat 
at high enough $Q^2$. The new experimental data however 
would rather be closer to $Q F_2 /F_1$. In Fig.~\ref{fig:ratios} 
both the ratio $Q^2 F_2/F_1$ and $G_{Ep}/G_{Mp}$ are depicted 
for the three different forms as compared to the new data. 
As before (this is nothing more than a different view at 
the same information contained in $G_{Ep}$ and $G_{Mp}$) 
front form gets closer to the high-$Q^2$ tendency of the data. 

Finally the values for the magnetic moments and proton 
charge radii are given in Table~\ref{tab:zeroq}. Instant 
and Front form get values for the magnetic moments which 
are in close agreement with the experimental data, 
while point form underestimates the value by 10 \%. 

\begin{table}[b]
\caption{Values of the form factors at $Q^2\rightarrow 0$ 
together with the proton charge radius.\label{tab:zeroq}}
\begin{tabular}{lcccc}
                &  Instant &  Point  &  Front & EXP \\  
\hline
 $G_{Mp}(0)$    &   2.7    &  2.5   & 2.8   & 2.793~\protect\cite{PDG}  \\
 $G_{Mn}(0)$    &$-$1.8    &$-$1.6  & $-$1.7  & $-$1.913~\protect\cite{PDG}\\
 $r_{cp}$(fm)   &   0.89   & 0.84  & 0.85  & 0.87~\protect\cite{PDG}\\
\end{tabular}
\end{table}

\subsection{Electromagnetic form factors of the neutron: 
mixed symmetry components}

\begin{figure}[t]
\vspace{20pt!}
\includegraphics[width=0.48\textwidth]{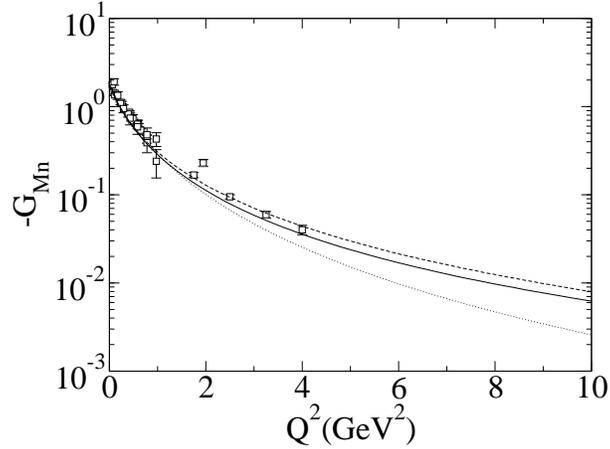}
\caption{Magnetic form factor of the neutron. Similar description as 
in Fig.~\ref{fig:gmp}.
\label{fig:gmn}}
\end{figure}

The neutron magnetic form factor comes out mostly in agreement with the 
experimental data as can be seen in Fig.~\ref{fig:gmn}. Similar features 
as for the proton case are found, instant and front forms provide similar 
descriptions in the considered $Q^2$ domain while point form tends to predict 
a lower value at high $Q^2$.

The neutron magnetic moment is given in table~\ref{tab:zeroq}. The values 
are in all cases smaller (in magnitude) than the experimental value, with 
point form giving the poorest number. In front form, the value is better 
but is still 10\% off. This is similar to what was found in Ref.~\cite{Chung}. 
There the possibility of anomalous magnetic moments for the quarks is also 
explored, and with that extra freedom the magnetic moments of both 
neutron and proton are reproduced precisely. It may be worth noting that 
in front form it is not possible, without anomalous magnetic moments, to 
get $G_{Mp}+G_{Mn} \leq 1$ as happens experimentally~\cite{coesterpriv}.

\begin{figure}[b]
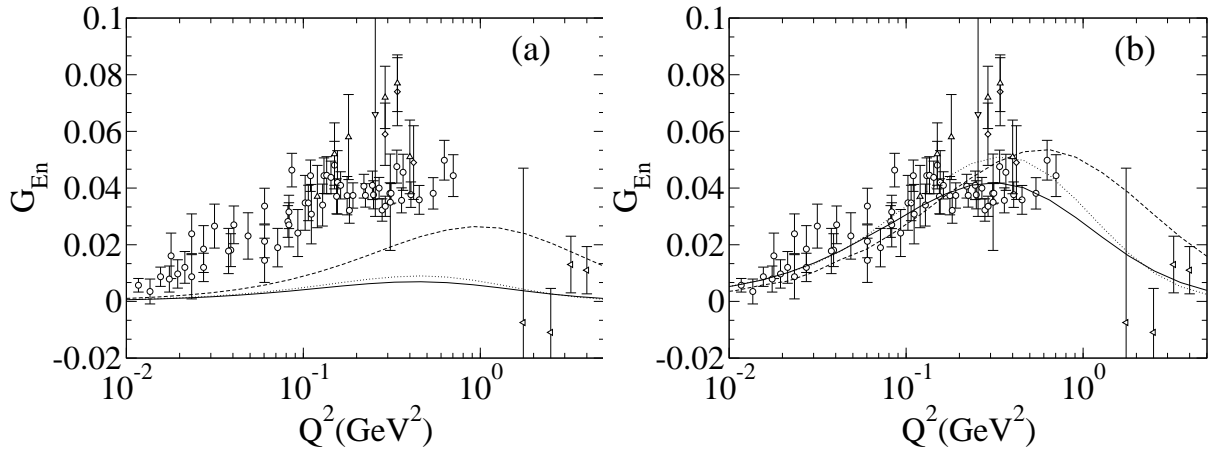

\vspace{20pt}
\includegraphics[width=0.48\textwidth]{fig_bjd_8a.eps}
\includegraphics[width=0.48\textwidth]{fig_bjd_8b.eps}
\caption{Electric form factor of the neutron. Left, symmetric s-wave wave 
function, right, with some admixture of mixed symmetry s-state. For curves 
description see caption of Fig.~\ref{fig:gmp}.
\label{fig:gen1}}
\end{figure}

In a symmetric non-relativistic quark model the electric form factor of the 
neutron is zero. Relativistic effects, which could in a sense deform the 
original symmetric shape in the rest frame, produce already non-zero values 
and a non-zero charge radius with the experimental sign. This is clear in 
Fig.~\ref{fig:gen1}, there the electric neutron form factor is plotted in 
all forms of kinematics. Noticeably, although non-zero, the values obtained 
with point and instant form are an order of magnitude lower than the 
experimental ones. The front form one is better, but still off the experimental 
data. 

The point form result differs with what was already known from 
Ref.~\cite{boffi}. The only difference between both calculations 
being the wave functions employed, in our case simple 2 parametric 
symmetric wave functions, while in their case complicated wave functions 
obtained from the solution of their quark-quark hamiltonian. We noted 
the fact that in their wave functions there was a small admixture of 
mixed-symmetry s-state whose origin was the interaction between quarks. 
Thus, a phenomenological admixture was included in our wave function 
as described in section II. The results obtained with a 1\% admixture 
are given in Fig.~\ref{fig:gen1}. The small admixture completely resolves the 
discrepancy in all the forms. The physical origin of such admixture must 
be sought in the quark-quark interaction.

\section{$N-\Delta$ transition in relativistic quark models}

The $N-\Delta$ electromagnetic transition is closely related to the 
presence of D-state components on both the $N$ and $\Delta$ rest frame 
wave functions. A vast number of experiments have been dedicated to 
explore this transition and to extract model independent data for this 
reaction~\cite{papanicolas,cole}. 

The physical questions at hand are many, first, it is an appropriate place 
to study the effect of the so-called pion cloud~\cite{satolee}, which in 
the quark model picture would correspond to exploring the effects of 
including more fock space configurations in the wave function, e.g. 
$|qqqq\bar{q}\rangle$~\cite{RiskaLi,ZouRiska}.

Second, this transition is sensitive to the presence of $L>1$ components on 
the $N$ and $\Delta$ wave function. The presence of such components is what is 
pursued when we talk about ``Shape of Hadrons'' (which is the title of this 
workshop). Several authors have discussed the issue of shape in the quark model 
framework in the recent years~\cite{Miller,Gross}, although it seems at times 
that there is no clear meaning to the word ``shape''. In the work of Miller~\cite{Miller} 
single quark spin dependent distribution functions are plotted as a proof of 
the existence of multiple shapes in the proton, in the work of Gross and 
Agbakpe~\cite{Gross}, on the contrary, it is stated explicitly that even the 
presence of D-state components on the proton would still render a symmetric 
charge distribution. 

Formally the $N-\Delta$ transition, can be characterized by the following set 
of form factors, 
\beq
\Gamma^\mu_\nu (P,Q)= \sum_i^3 {\cal K}_\nu^{i,\mu}(P,Q)G_i(Q^2)\, ,
\eeq
with,
\beqa
 {\cal K}^{1,\nu\mu}(P,Q) &=& {Q^\nu \gamma^\mu - (\gamma\cdot Q) 
g^{\nu\mu}\over \sqrt{Q^2}}\sqrt{M^* M} 
\gamma_5\, , \nonumber \\
 {\cal K}^{2,\nu\mu}(P,Q) &=& {Q^\nu P^\mu - (P\cdot Q) g^{\nu\mu
}\over \sqrt{Q^2}} \;\gamma_5\, , \nonumber \\
 {\cal K}^{3,\nu\mu}(P,Q) &=& {Q^\nu Q^\mu - Q^2
 g^{\nu\mu}\over Q^2 } M^* \;\gamma_5 \, ,
\eeqa
where $M$ and $M^*$ are the nucleon and resonance masses respectively.
The Sachs like magnetic dipole, electric quadrupole and Coulomb form 
factors are defined as in Ref.~\cite{Scad},
\beq
\Gamma^\mu_\nu (P,Q)= 
G_M^*(Q^2) {\cal K}_\nu^{M,\mu}(P,Q)
+G_E^*(Q^2) {\cal K}_\nu^{E,\mu}(P,Q)
+G_C^*(Q^2) {\cal K}_\nu^{C,\mu}(P,Q) \,.
\eeq

The relation between the three form factors $G_1, G_2$ and $G_3$ and the 
corresponding electric, magnetic and Coulomb form factors are~\cite{Scad}:
\beqa
 G_E^*&=&{M\over 3(M^*+M)} \left[{M^{*2}-M^2-Q^2\over M^*}
{\sqrt{M^* M} \over Q} G_1
+{M^{*2}-M^2\over Q}G_2-2M^*G_3 \right]\, ,\nonumber \\
 G_M^*&=&{M\over 3(M^*+M)}\left[{(3M^*+M)(M^*+M)+Q^2\over M^*}
{\sqrt{M^* M} \over Q}G_1 
   + {M^{*2}-M^2\over Q} G_2-2 M^* G_3\right]\, ,
\nonumber \\
  G_C^*&=&{2M\over 3(M^*+M)}\left[{ 2 M^*\sqrt{M^* M} \over Q} G_1 +
{3M^{*2}+M^2+Q^2 \over 2 Q} G_2 
+ {M^{*2}-M^2-Q^2\over Q^2}  M^*G_3\right]  \, .
\label{oursEMC}
\eeqa
In instant and point form kinematics the relation between the different 
spin state matrix elements of the electromagnetic current and the form 
factors, $G_j$, is finally:
\begin{eqnarray}
I_{\thalf,\ohalf}^1
&=&\left[ {M^* +M\over \sqrt{ Q^2}} G_1 + 
{M^{*2}-M^2\over 2\sqrt{Q^2 M M^*}}G_2
-{\sqrt{M^*\over M} }G_3\right]
  {Q_3\over 2 \sqrt{E(M+E)}} \, ,
\nonumber\\
I_{\ohalf,-\ohalf}^1
&=&-{\sqrt{3}\over 6}\left[{M^*+M\over \sqrt{ Q^2}}G_1
+ {M^{*2}-M^2\over 2\sqrt{ Q^2 M M^*}}G_2-
{\sqrt{M^*\over M}}G_3\right]
  {Q_3\over \sqrt{E(M+E)}}  
+{\sqrt{3}\over 3}
{Q_3\over \sqrt{Q^2}}{M+E\over \sqrt{E(M+E)}}G_1\, ,
\label{Gs}  \nonumber \\
I_{\ohalf,\ohalf}^0
&=&- {\sqrt{3}\over 3}
\left[{Q_3\over \sqrt{Q^2}}G_1
+{Q_3\over \sqrt{Q^2M M^*}}{E+M^*\over 2} G_2
+ {Q_3 Q_0 \over Q^2}\sqrt{M^*\over M} G_3 \right] 
 {Q_3 \over \sqrt{ E (M+E)} } \, . 
\end{eqnarray}
Here the 4-momentum transfer is taken as $Q=\{Q^0,0,0,Q_3\}$, 
with $I_{{j_\Delta},{j_N}}^m=\langle j_\Delta, 
P_\Delta\vert I_m(0)\vert P_N,j_N \rangle$,
and
\beq
Q^0 = -{P^* \cdot Q \over M^*} =  {M^{*2} - M^2 -Q^2\over 2 M^*}\,, 
\qquad  Q_3=\sqrt{Q^2+Q^{0\,2}} \, .
\eeq

The $E2/M1$ and $C2/M1$ ratios for the $\Delta-N$ 
transition are defined 
as:
\beqa
&&R_{EM}\equiv {E_2 \over M_1} \equiv - {  G_E^* \over G_M^*}\, , 
\nonumber \\
&& R_{SM}\equiv{C_2 \over M_1} \equiv  {|\vec q|\over 2 M^*}
 {  G_C^* \over G_M^*}\,.
\eeqa
Here $2M^* |\vec q|= \left([Q^2 + (M^*-M)^2][Q^2 + (M^*+M)^2]\right)^{1/2}$.

\subsection{Difficulties in the front form formulation}

In the application of front form kinematics it has been conventional to adopt 
a reference frame in which $Q^+=0$. However, as explored in detail in 
Ref.~\cite{dstate} even if one works in such frame the equivalent to the 
angular condition is widely violated. That is, in front form there are four 
spin amplitudes which are linear combinations of only three form factors, 
$G_i$, therefore they must be linearly dependent. This has also been studied 
by Cardarelli {\it et al.}~\cite{simula,simula2} for the same process, 
showing that in the single quark current approximation being used the 
angular condition cannot be fulfilled, implying that relativity is not 
correctly implemented. The reader is referred to those references for a deeper 
understanding of this point and we proceed to explore point and 
instant form calculations in our adopted single quark current approximation.

\subsection{Results for instant and point form}

Ratios of these form factors are commonly named $R_{EM}$ and $R_{SM}$. There 
are definite pQCD predictions for them, at $Q^2 \to \infty$, $R_{EM}\to 1$. The
non-relativistic quark model predictions at $Q^2=0$ yield $R_{EM}=R_{SM}=0$. 
Already 20 years ago, Ref.~\cite{Weber} studied the variation of $R_{EM}$ 
in a relativistic quark model (front form) with symmetric rest frame 
wave functions. They found that $R_{EM}$ would indeed become non-zero but 
extremely small, lower than $0.2 \%$. 

\begin{figure}[t!]
\vspace{25pt} 
\includegraphics[width=0.65\textwidth]{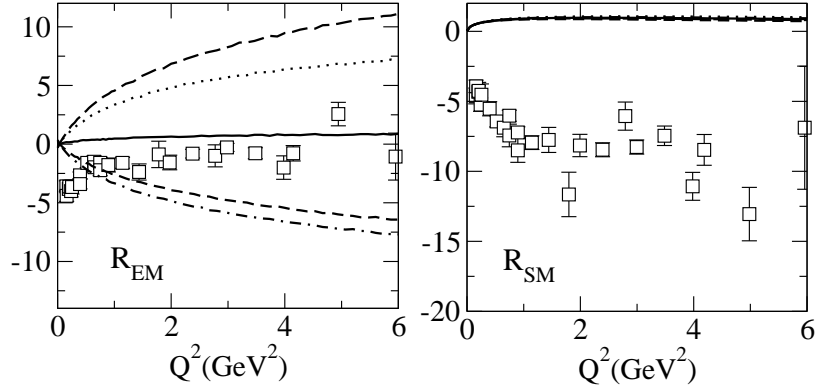}
\caption{$R_{EM}$ and $R_{SM}$ ratios in percent obtained 
in instant form. Solid, dotted, dashed, long-dashed and dot-dashed 
lines stand for $(b_N,b_\Delta)=$ (0,0), (0.2,0), ($-$0.2,0)
(0,0.2) and (0,$-$0.2) respectively. The experimental data 
are from the compilation of Ref.~\cite{Harry}.
\label{emratioI}}
\vspace{10pt}
\end{figure}

Here their result is extended to the other forms of kinematics making 
use of the ground state wave functions found in the elastic case. Then, the 
simpler way to phenomenologically obtain structure in those ratios is by 
considering a D-state configuration in the rest frame wave function of both 
the $N$ and the $\Delta$ as shown above. To explore all the possibilities 
the following wave functions for the $N$ and $\Delta$ are considered:
\beqa
\phi_N &=& a_N \phi_S + b_N \phi_D \nonumber \\
\phi_D &=& a_\Delta \phi_S + b_\Delta \phi_D \nonumber \\
\eeqa
with $|N_S|^2+|N_D|^2=1=D_S|^2+|D_D|^2$. In this exploratory study we are 
interested in the effect of $D-$state configurations in the wave functions. 
The following cases are considered:  ($b_N$,$b_\Delta$)= (0,0), (0.2,0), 
($-$0.2,0), (0,0.2) and (0,$-$0.2).

\begin{figure}[t]
\vspace{20pt!} 
\includegraphics[width=0.65\textwidth]{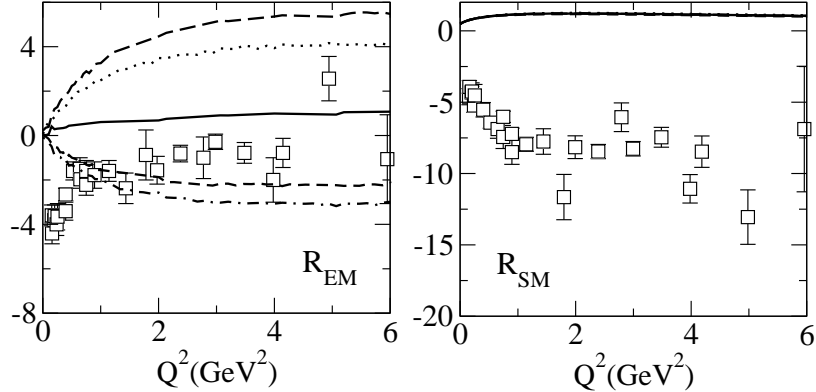}
\caption{$R_{EM}$ and $R_{SM}$ ratios in percent obtained 
in point form. Solid, dotted, dashed, long-dashed and dot-dashed 
lines stand for $(b_N,b_\Delta)=$ (0,0), (0.2,0), ($-$0.2,0)
(0,0.2) and (0,$-$0.2) respectively. The experimental data 
are from the compilation of Ref.~\cite{Harry}.\label{emratioP}}
\label{remp}
\vspace{10pt} 
\end{figure}

In figures~\ref{emratioI}
and~\ref{emratioP} plots for the obtained results 
for instant and point form kinematics are given. The main features 
of the results are shared by both implementations of relativity, 
essentially when no $D-$state configuration is considered both 
the $R_{EM}$ and $R_{SM}$ ratios are small and quite different 
from the experimental data. The inclusion of a $D-$state 
configuration affects mainly the electric form factor and thus 
shows up abruptly in the $R_{EM}$ ratio. The relative sign of the $D-$state 
configuration is relevant, it is found that a negative sign would be in line 
with the experimental data. This is so irrespective of whether it is the 
$N$ or the $\Delta$ which contains the D-state configuration.

The situation is completely different for the $R_{SM}$ ratio, 
there, the inclusion of such configuration does not give any 
seizable effect in either form. 

Concerning the values at $Q^2=0$ it becomes apparent from the 
figures that although they are not zero, due to the relativistic 
nature of the calculations, they are far from the few percent 
experimental values. Thus, the findings of Ref.~\cite{Weber} 
are confirmed also for instant and point form.  

\section{Summary and conclusion}

Nucleons and low lying resonances in the energy domain of a few GeV are 
the perfect laboratory to study low energy QCD. There is a rich variety 
of resonances, meson-baryon states and maybe more exotic structures which 
coexist and which cannot yet be understood from the successful theory. 
The use of lattice QCD is a very promising one, and a great effort is 
being devoted to its implementation. At the same time, the effective field 
theory of the $N-\Delta$ has also been presented 
(see Ref.~\cite{pascalutsa,gail}) 
and in its range of validity tends to agree with the experimental data. 
This effective theory, on the contrary, is only valid on a very narrow range. 

Quark models, on the other hand, provide a organizational tool which allows 
to explore several relevant issues in strong physics. Here the way a quark 
model, which uses a simple spectral representation of the 
mass operator, is able to accommodate to a great extend the experimental 
form factors of the nucleon has been explained. Further refinements of 
the model in each of the forms could of course be pursued and a finer 
agreement with the data could most likely be achieved. An improvement 
which has been studied is the inclusion of anomalous magnetic moments 
for the constituent quarks, this of course would help resolve some 
of the discrepancies with the 
experimental data. On the other hand, up to now mostly three quark 
configurations have been considered when exploring the electromagnetic 
structure of hadrons and also in most studies of the baryon spectrum in 
quark models. The inclusion of higher components, both from a phenomenological 
point of view, and also from the solution of dynamical equations for the 
multiquark system needs to be further investigated. Small percentages of 
such higher configurations could play important roles in photoproduction 
and also in meson decays of resonances~\cite{RiskaLi}. This is to say, 
all the sensible configurations need to be exhausted before actually 
going into more complicated microscopic assumptions, such as the 
anomalous magnetic moments. In fact, the consideration of higher fock 
states is the natural step in the quark model in view of the recent 
dynamical studies which explain, to a certain extend, some resonances 
as molecular meson-baryon states. As an example, the dynamical model 
of Sato and Lee~\cite{satolee}, is able to reproduce the quotients $R_{EM}$ and 
$R_{SM}$, being able to track down this success to their account of the 
so-called meson cloud effects.

Our study was aimed at exploring both the consequences of the different 
implementations of relativity and at the same time of the incorporation of 
non-S wave configurations on the wave functions. 

With that in mind, the main learnings are: 
\begin{itemize}
\item The three forms of kinematics are able to describe the elastic form 
factors of the nucleon in a very reasonable way after fitting 2 parameters 
in the wave function plus the constituent quark mass. The fit is only 
constraint by the magnetic form factor of the proton.
The higher tail of the form factors was however systematically underestimated
by the point form calculation. The neutron electric form factor requires 
the inclussion of a small $\approx 1\%$ mixed-symmetry component on the 
rest frame wave function. 

\item The only qualitative difference between the three forms 
appears in the presence of a node in the electric form factor 
of the proton. This node only appears, with the pointless quarks 
used here, in the front form calculation. The inclusion of mixed-symmetry 
components or $D-$state configurations would not change this picture. The 
node is in better agreement with the expected behavior after the recent 
JLAB measurements. The presence of nodes in front form form factors occurs 
also in the case of the rho and kaon form factors, 
see Ref.~\cite{junheplb,junheepja}.

\item The issue of shape, phrased as importance of D-state components, 
has been addressed for the $N-\Delta$ transition. There the results 
are both encouraging and not so. First, none of the forms is able to 
account for the structure present in the quotients, $R_{EM}$ and $R_{SM}$, 
even when a D-state configuration is considered. Consideration of other 
fock state components, exchange currents and possibly resorting to 
anomalous magnetic moments for the constituent quarks could help resolve 
the problem. Front form calculations using single quark currents are 
of no interest for this transition as they are not fully relativistic.  
\end{itemize}

%%%%%%%%%%%%%%%%%%%%%%%%%%%%%%%%%%%%%%%%%%%%%%%%
%% BACKMATTER
%%%%%%%%%%%%%%%%%%%%%%%%%%%%%%%%%%%%%%%%%%%%%%%%

\begin{theacknowledgments}
The author wants to thank D.O. Riska for a careful reading of the 
manuscript and for the on going collaboration which produced 
most of the results presented here. This presentation has been 
funded in part by the EU Integrated Infrastructure Initiative
Hadron Physics Project under contract number RII3-CT-2004-506078
\end{theacknowledgments}

%%%%%%%%%%%%%%%%%%%%%%%%%%%%%%%%%%%%%%%%%%%%%%%%
%% The bibliography can be prepared using the BibTeX program or
%% manually.
%%
%% The code below assumes that BibTeX is used.  If the bibliography is
%% produced without BibTeX comment out the following lines and see the
%% aipguide.pdf for further information.
%%
%% For your convenience a manually coded example is appended
%% after the \end{document}
%%%%%%%%%%%%%%%%%%%%%%%%%%%%%%%%%%%%%%%%%%%%%%%%

%%%%%%%%%%%%%%%%%%%%%%%%%%%%%%%%%%%%%%%%%%%%%%%%
%% You may have to change the BibTeX style below, depending on your
%% setup or preferences.
%%
%%
%% For The AIP proceedings layouts use either
%%%%%%%%%%%%%%%%%%%%%%%%%%%%%%%%%%%%%%%%%%%%

\bibliographystyle{aipproc}   % if natbib is available
%\bibliographystyle{aipprocl} % if natbib is missing

%%%%%%%%%%%%%%%%%%%%%%%%%%%%%%%%%%%%%%%%%%%
%% You probably want to use your own bibtex database here
%%%%%%%%%%%%%%%%%%%%%%%%%%%%%%%%%%%%%%%%%%%
\bibliography{sample}

%%%%%%%%%%%%%%%%%%%%%%%%%%%%%%%%%%%%%%%%%%%
%% Just a reminder that you may have to run bibtex
%% All of it up to \end{document} can be removed
%% if you don't like the warning.
%%%%%%%%%%%%%%%%%%%%%%%%%%%%%%%%%%%%%%%%%%%
\IfFileExists{\jobname.bbl}{}
 {\typeout{}
  \typeout{******************************************}
  \typeout{** Please run "bibtex \jobname" to optain}
  \typeout{** the bibliography and then re-run LaTeX}
  \typeout{** twice to fix the references!}
  \typeout{******************************************}
  \typeout{}
 }

\end{document}